\pdfoutput=1
\documentclass[jkps,fleqn,showkeys,twocolumn]{revtex4-2}
\usepackage[pdftex]{graphicx}
\usepackage{amssymb}
\usepackage{amsmath}
\usepackage{bm}

\usepackage[version=4]{mhchem}                  
\usepackage{verbatim}                           
\usepackage{booktabs}                           
\newcommand*{\saa}{$\mathrm{^{(a)}}$}           
\newcommand*{\sbb}{$\mathrm{^{(b)}}$}           
\newcommand*{\scc}{$\mathrm{^{(c)}}$}           
\def\finalVersion{}                 
\ifdefined\finalVersion             
\newcommand*{\added}[1]{{#1}}       
\newcommand*{\deleted}[1]{}         
\newcommand*{\removed}[1]{}         
\newcommand*{\replaced}[2]{#2}      
\else                               
\usepackage{xcolor}                 
\usepackage[normalem]{ulem}         
\newcommand*{\added}[1]{\textcolor{blue}{#1}}
\newcommand*{\deleted}[1]{\textcolor{red}{\sout{#1}}}
\newcommand*{\removed}[1]{\textcolor{red}{\sout{#1}}}
\newcommand*{\replaced}[2]{\textcolor{red}{\deleted{#1}\added{#2}}}
\fi

\begin{document}
\setcounter{page}{0}
\title[]{Molecular-Beam Spectroscopy with an Infinite Interferometer: Spectroscopic Resolution and Accuracy}
\author{Thomas Schultz}
\email{schultz@unist.ac.kr}
\author{In Heo}
\author{Jong Chan Lee}
\author{Beg\"um Rukiye \"Ozer}
\affiliation{Department of Chemistry, Ulsan National Institute of Science and Technology (UNIST),
44919, Korea}

\date[]{Received September 20, 2022}

\begin{abstract}
An interferometer with effectively infinite maximum optical path difference removes the dominant resolution limitation for interferometric spectroscopy. We present mass-correlated rotational Raman spectra that represent the world's highest resolution scanned interferometric \replaced{spectra}{data} and discuss the \added{current and} expected \added{future} limitations in achievable spectroscopic performance.
\end{abstract}


\maketitle

\section{Introduction}

The resolution of scanned interferometric spectroscopy is usually limited by the maximum optical path difference (MOPD) of the interferometer. As part of our development of mass-correlated rotational alignment spectroscopy (CRASY),\cite{Schroter2011} we constructed an interferometer with an effectively infinite MOPD,\cite{Schroter2018} utilizing the discrete pulse train emitted from a femtosecond laser oscillator. Here we present the highest resolution data measured to-date and discuss the resolution-limiting factors in our molecular beam experiments. Discussed concepts are equally applicable to other types of interferometric or time-domain spectroscopy that rely on the scanning of a spatial path difference or a temporal delay range.

CRASY is a type  of rotational coherence spectroscopy (RCS)\cite{Frey2011} and measures rotational Raman spectra in the time-domain by scanning the path difference between two interferometer arms. The resulting data is useful for the structural characterization of non-dipolar molecules\cite{Schroter2011,Schroter2015,Heo2022a,Heo2022b} that are inaccessible to Fourier-transform microwave Spectroscopy (FTMW).\cite{Grabow2011} By extending the scanned interferometer length, our work obtained order-of-magnitude improved resolution\cite{Schroter2018,Lee2019} as compared to preceding RCS experiments or Fourier-transform infrared spectroscopy (FTIR).\cite{Albert2011}

The energy resolution ($\Delta E$) of spectroscopic experiments is fundamentally limited by the observation time $\Delta t$, i.e., the time over which the investigating particles (usually photons) and investigated molecules interact. The time-frequency formulation of Heisenberg's uncertainty principle states this fundamental resolution limit as $\Delta E \cdot \Delta t = \hbar/2$. Independent of the experiment, the effective observation time is always limited by the coherence time (or lifetime) of the observed quantum states. This leads to lifetime-broadening of observed spectral lines, either due to an intrinsically limited lifetime of the observed states or due to interactions with an environment, e.g., through molecular collisions. When lifetime broadening is small, the effective resolution will be limited by the constraints of the spectroscopic experiment.

In frequency domain spectroscopy, the coherence length of the interacting photons limits the effective observation time. The situation is fundamentally different in interferometric Fourier-transform spectroscopy, where the observation time is limited by the MOPD \replaced{that can be achieved within}{of} the interferometer. The non-apodized full-width-at-half-maximum (FWHM) resolution limit in FTIR is given as $\Delta \widetilde{\nu}^{FWHM} = 0.61 \cdot \mathrm{d^{-1}_{MOPD}}$.\cite{Albert2011} The highest-resolution interferometer described in the literature features a MOPD of 11.7~m,\cite{Albert2018} which corresponds to a delay range of $t_\textrm{max} = 39$ ns and a non-apodized resolution limit of 15.6 MHz FWHM.

RCS is based on the impulsive excitation and probing of rotational coherence with ultrafast (femtosecond or picosecond) laser pulses. In our CRASY variant of RCS, the rotationally excited molecules are probed by resonant multi-photon photoionization and rotational coherence is observed as interferometric signal modulation \replaced{of}{in} resulting ion signals. CRASY therefore correlates rotational spectra with observed ion masses and thereby facilitates the assignment of signals in heterogeneous samples. As in FTIR, or other scanned interferometric spectroscopies, RCS experiments scan the optical path difference in an interferometer and shows a resolution limited by the MOPD.

\section{Infinite Interferometer Design for CRASY}

\begin{figure}[ht]
	 \includegraphics[width=240pt]{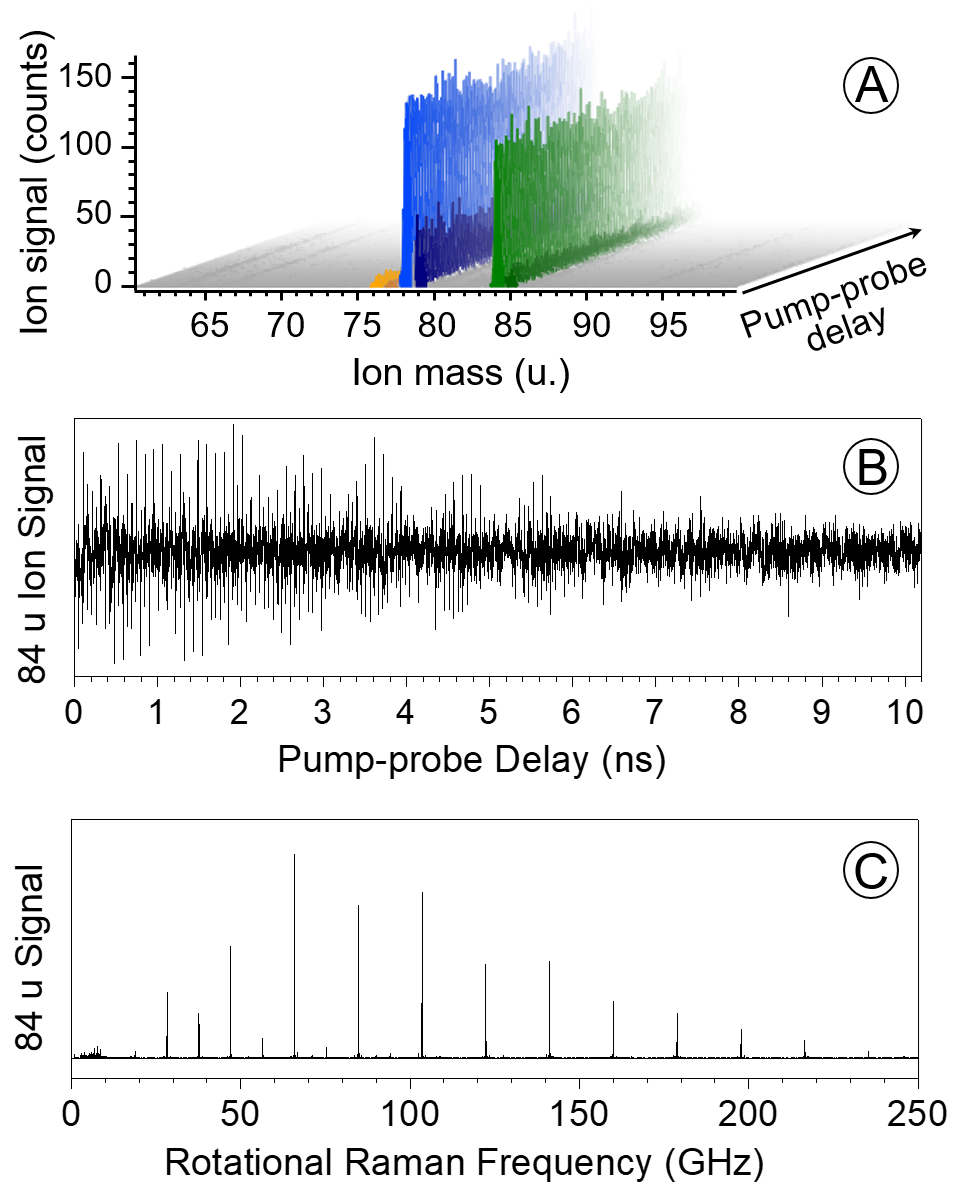}
	\caption{Mass-CRASY data from a 50 ns delay scan of a sample containing benzene (mass 78 u, blue), perdeuterated benzene (mass 84 u, green), carbon disulfide (mass 76 u, orange), and naturally occurring heavy isotopologues (darker colors). (A) Delay dependent ion signals show significant signal modulation due to interference of coherently excited rotational states. (B) A section of the signal modulation trace for mass 84 u. (C) Rotational Raman spectrum obtained by Fourier-transformation of the trace shown in (B).}
	\label{CRASY_data_example}
\end{figure}

To obtain mass-CRASY data, ion signals were detected in a time-of-flight mass spectrometer as function of the scanned delay between alignment and ionization laser pulses. The experimental details for mass-CRASY measurements were described previously \cite{Schroter2015,Schroter2018,Lee2019,Ozer2020,Lee2021} and here we focus on the interferometer design used to scan extended optical path differences. Focussed alignment pulses with 800 nm wavelength, $\leq$2 ps pulse duration and 100 $\mu$J-level pulse power created a coherent rotational wavepacket by impulsive Raman excitation. Alignment, in this context, denotes the transient molecular alignment that is commonly observed upon excitation of a coherent wavepacket.\cite{Stapelfeldt2004} Ionization pulses with 200 nm or 266 nm wavelength, 45 fs pulse duration and few-$\mu$J pulse power photoionized molecules by two-photon resonant photoionization. Ion signals showed temporal signal modulations due to the interference of the coherent rotational states in the probe step, as depicted in Fig\ \ref{CRASY_data_example} (A) and (B). Fourier-transformation of these signal modulations reveal the spectrum of the coherent wavepacket, as shown in Fig.\ \ref{CRASY_data_example} (C).

The interferometer for high-resolution spectroscopy should have the longest possible MOPD, combined with a small step size and high positioning accuracy. As described above, the achievable spectroscopic resolution is directly proportional to the MOPD and, as described by the Shannon-Nyquist theorem,\cite{Shannon1949} the spectroscopic range $\nu_\mathrm{max}$ is inversely proportional to the sampling step size ($\nu_\mathrm{max} = 1/(2 \cdot t_{step\:size}$)). CRASY is performed on cold molecular beams with beam temperatures below 10 K and a 0.5 ps to 5 ps steps size (maximum spectroscopic range of $\nu_\mathrm{max} =$ 0.1 THz to 1 THz) is sufficient to resolve the complete thermally occupied set of rotational states. The positioning accuracy should remain well-below the scanned step size to avoid a degradation of \added{the} spectroscopic resolution.

Interferometers used for FTIR, RCS, and other types of scanned interferometer spectroscopy are based on opto-mechanical delay stages, i.e., moving mirrors in one interferometer arm to change the optical path length. The largest interferometers can be found at national synchrotron facilities but are not practical within the restricted space and budget of University-based laboratory research. Instead, our interferometer combined electronic and opto-mechanical delays to achieve an infinite effective delay range within a compact and affordable interferometer design.

\begin{figure}[ht]
	 \includegraphics[width=240pt]{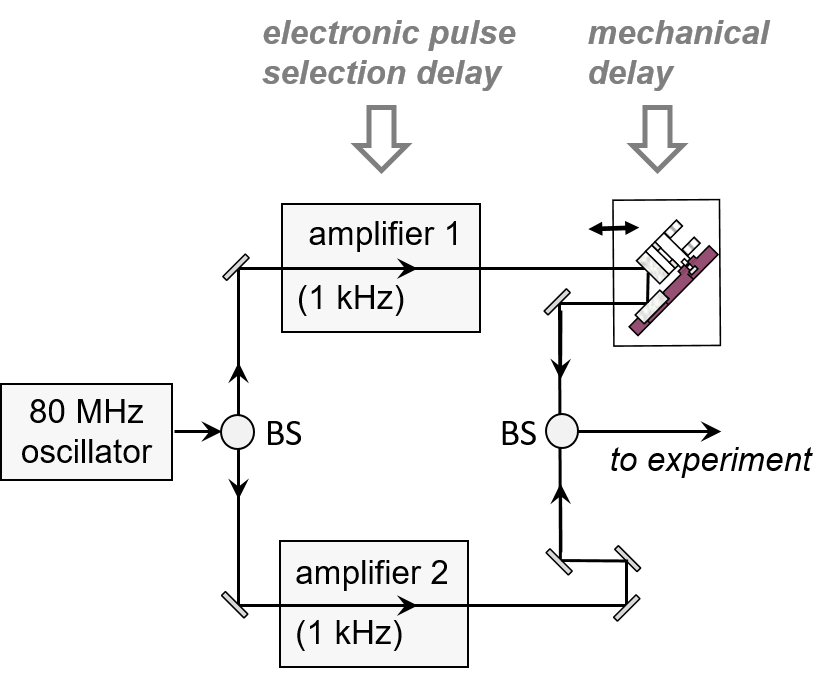}
	\caption{Schematic depiction of the infinite interferometer design. Pulses from a femtosecond laser oscillator are split and recombined with beam splitters (BS) and amplified in two separate amplifiers. Electronic selection of different oscillator pulses for amplifier 1 and 2 introduce\replaced{s}{d} discrete pulse delays in multiples of 12.5 ns. An opto-mechanical delay stage add\replaced{s}{ed} additional delays of 0--16 ns with picosecond step size and femtosecond accuracy. }
	\label{InterferometerDesign}
\end{figure}

A schematic representation the infinite interferometer is shown in Fig.\ \ref{InterferometerDesign}. The mechanical delay was based on a 30-cm Physik Instrumente, MD-531 motorized stage. The optical beam path across the stage was 16-times folded to obtain a MOPD of 4.8 m (16 ns). Longer delays were achieved by electronic pulse selection of oscillator pulses that were amplified in two separate regenerative Ti:Sa amplifiers, forming the two arms of the interferometer. The repetition rate of the laser oscillator (Coherent Vitara) was a 80 MHz and selection of subsequent pulses added discrete 12.5 ns delays in the second interferometer arm. Note that the timing accuracy of the electronic delays is governed by the stability of the oscillator repetition rate and not the accuracy of the electronic delay generator.

The pulse selection delay was controlled via an electronic delay generator (SRS-DG535) and allowed to extend the delay range to quasi-arbitrary values. The amplifiers were operating at 1-kHz repetition rate and probe pulses delayed by more than 1 ms therefore arrive after a subsequent pump pulse. The molecular beam velocity in our experiments is in the range of 1000 m/s and molecules travel \replaced{decimeter}{meter} distances within milliseconds. Experiments with $>1$ ms delays can therefore rely on spatial discrimination of the excited molecules. Therefore, for all practical purposes, our set-up represents an interferometer with infinite MOPD and the achievable spectroscopic resolution is no longer limited by the size of the interferometer, but \replaced{rather by}{by other experimental limitations, such as }the ability to track the molecular beam.

\section{High-Resolution CRASY Data}

\begin{figure}[ht]
  \includegraphics[width=240pt]{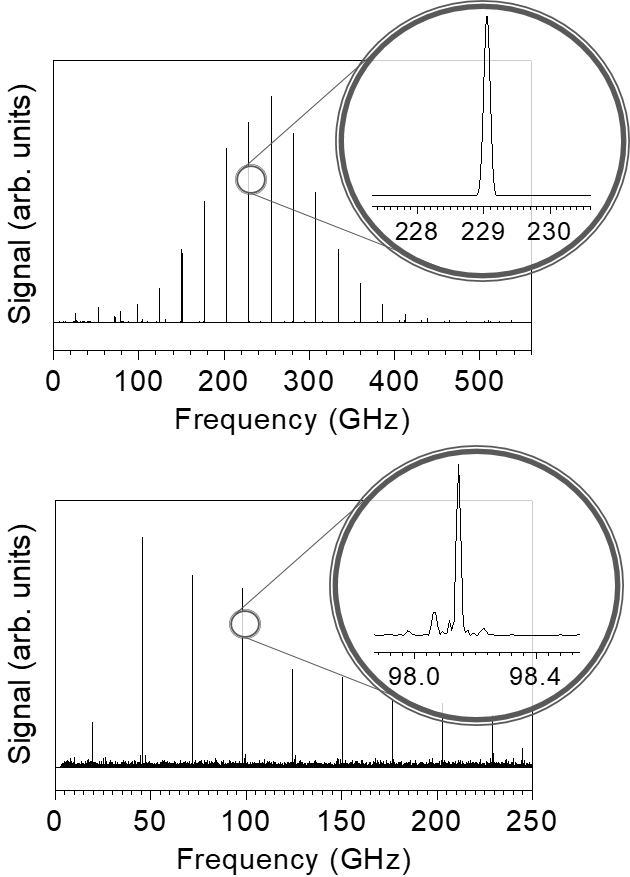}
	\caption{Rotational Raman spectra for \ce{CS2} from CRASY data-sets. (Top) Spectrum obtained by scanning a 16 ns delay range (4.8 m MOPD) with an opto-mechanical delay stage. (Bottom) Spectrum obtained by scanning a 56 ns delay range (equivalent to 16.7 m MOPD), using combined opto-mechanical and electronic delays. Enlarged insets reveal the effective resolution for selected transition lines.}
	\label{CS2_16_vs_50_ns}
\end{figure}

Fig.\ \ref{CS2_16_vs_50_ns} shows rotational Raman spectra for \ce{CS2} and illustrates the progress achieved by combining electronic and opto-mechanical delays. The maximal opto-mechanical delay range of 16 ns was sufficient to obtain an effective resolution of 60~MHz FWHM, as shown in Fig.\ \ref{CS2_16_vs_50_ns} (Top). The effective resolution remained below the non-apodized resolution limit of 38 MHz because the mechanical delay stage was not perfectly flat, leading to a loss of signal when the stage approached either end of the delay range. A combined opto-mechanical and electronic delay range of 16.7 m (50 ns) gave a greatly enhanced effective resolution of 17.5 MHz FWHM, near the resolution limit of 16.7 MHz, as shown in Fig.\ \ref{CS2_16_vs_50_ns} (Bottom). The sample \replaced{for the latter spectrum}{used for the latter measurement} contained only trace amounts of \ce{CS2} and the spectrum therefore had a lower signal-to-noise ratio \added{(SNR)}.

The measurement time required to scan large delays scales linearly with the scan range and the collection of mass spectra for a large number of alignment-ionization delays was time-consuming and created exceedingly large data-sets. The data shown in Fig.\ \ref{CS2_16_vs_50_ns} (Bottom) was obtained with delay scan range of 50 ns and a 2 ps step-size and therefore required the accumulation of 25\,000 mass spectra. Data was acquired with 500 Hz repetition rate and ion signals for 1000 laser shots were accumulated for each mass spectrum. The resulting measurement time was almost 14 hours. Each time-of-flight mass spectrum contained 400\,000 point\added{s} and the acquired raw data quantity corresponds to 10 Gb. It is readily apparent that the brute-force extension of the scanned delay range will lead to impractical requirements in terms of measurement time and \deleted{for} data storage.

We reduced the data quantity by lossless compression and the use of sparse data formats. Because mass spectroscopic data is highly discrete, we routinely achieved $>100$-fold in-memory compression with zlib compression algorithms. Fourier transform analysis is only possible on decompressed data, but downsampling of the mass axis and the conversion into sparse data formats facilitated the signal analysis.

\begin{figure}[ht]
  \includegraphics[width=240pt]{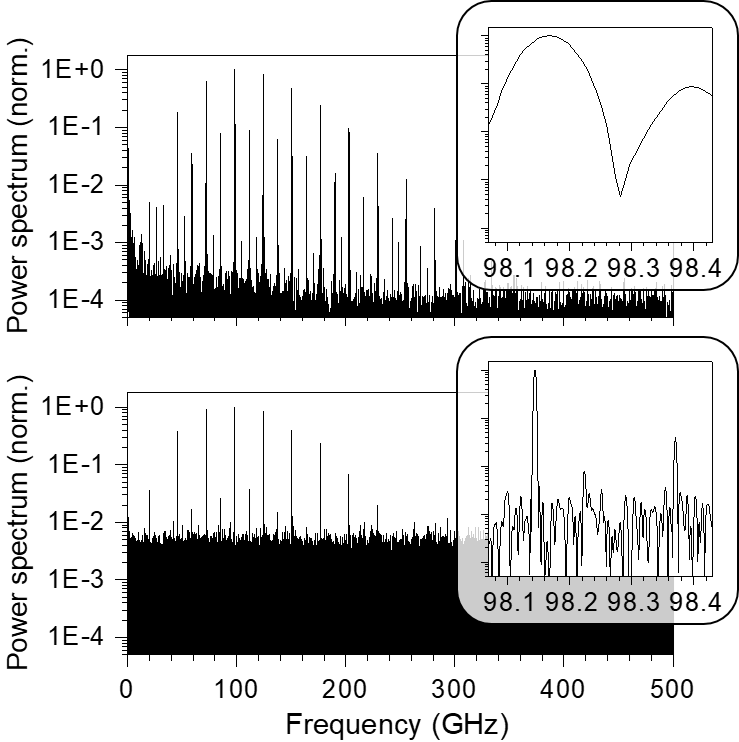}
	\caption{Rotational Raman spectra for \ce{CS2} obtained with continuous 1-ps sampling of a 15 ns delay (Top) and random sparse sampling of 17\,000 mass spectra along a 312 ns delay \added{axis }(5.5\% sampling rate, Bottom). Note the logarithmic scale of the ordinate.}
	\label{M76_full_and_sparse_sampling}
\end{figure}

Random sparse sampling was used to accelerate long delay scans, i.e., data was only measured for a randomly selected sub-set of delay\replaced{s along the}{ points along an} extended time axis. Different sparse sampling strategies were explored in the field of multi-dimensional NMR experiments,\cite{Hoch2014,Pelczer1991} and are discussed in more detail, below. Fig.\ \ref{M76_full_and_sparse_sampling} compares spectra from a fully-sampled and a sparsely-sampled measurement, with the latter collecting mass spectra only for 5.5\% of delays along an extended delay axis. The sparsely sampled data was acquired 2.5-times faster than the fully sampled data and improved the spectroscopic resolution by a factor 20. Sparse sampling added noise to the spectra, as readily apparent in the logarithmic representation depicted in Fig.\ \ref{M76_full_and_sparse_sampling}.

\begin{figure}[ht]
\centering
  \includegraphics[width=240pt]{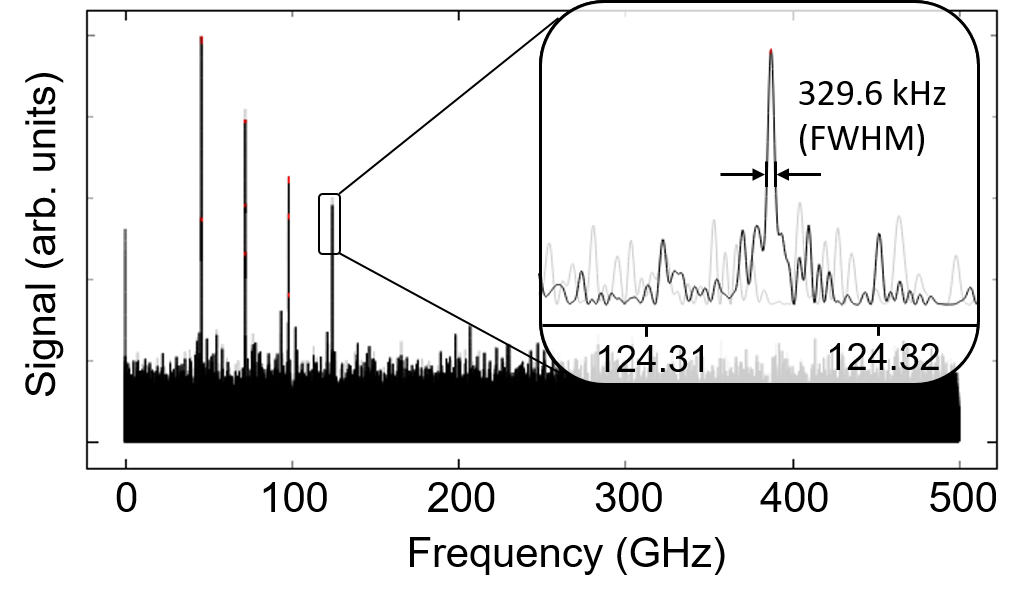}
  \caption{\small Highest resolution rotational Raman spectrum obtained with the mass-CRASY technique. The inset, with 30\,000-fold enlarged abscissa, reveals the 330 kHz FWHM \added{effective} resolution for the J=6--8 transition in \ce{CS2}. }
  \label{fig:kHz_resolution_Spectrum}
\end{figure}

The highest resolution spectrum measured to-date with the CRASY technique was based on a 10 $\mu$s scan of a benzene sample, containing residual \ce{CS2} in small concentration. Due to the limited size of our spectrometer window, tracking of the molecular beam was only achieved \replaced{up to}{for a delay range} $<3$ $\mu$s, reducing the achieved signal contrast and resolution. Fig.\ \ref{fig:kHz_resolution_Spectrum} show\added{s} signal for the \ce{CS2} mass channel, with an effective resolution of 330 kHz FWHM. This data represents the highest-resolution Fourier-transform interferometric spectrum in the world, representing a 50-fold improvement over the highest-resolution FTIR data in the literature\cite{Albert2011,Albert2015,Albert2018}. To comprehend the scale of this improvement, we invite the reader to visualize the \removed{large} 11.7 m MOPD interferometer used for the latter experiments (see Ref.\ \citenum{Albert2011b} for a photographic image) versus the km-scale MOPD achieved in our laboratory experiment.

Table \ref{tab:Spectroscopic_resolution} compares the resolution limit of various spectroscopic techniques that are used to characterize rotational spectra at high resolution. CRASY currently represents the highest-resolution method for rotational Raman spectroscopy and \added{, more generally,} for the investigation of non-dipolar molecules. Modern FTMW experiments reach a significantly higher resolution,\cite{Shipman2011} but can only be performed for dipolar species and only cover a spectral range of tens of GHz, more than one order-of-magnitude below the spectral coverage obtained with CRASY. In terms of resolving power ($\frac{\mathrm{spectral\:range}}{\mathrm{resolution}}$), CRASY is at parity with state-of-the-art FTMW experiments.

\begin{table}
 \caption{Resolution for common types of rotationally resolved spectroscopies.}
\begin{ruledtabular}
 \begin{tabular}{lc}
{Spectroscopic method}           &   {Resolution limit}\\
\colrule
Raman, single-mode laser\saa     & 	  1500 MHz    \cite{Weber2011}\\
Raman, FTIR\saa	                 &     300 MHz    \cite{Weber2011}\\
Raman, RCS\sbb                   &     150 MHz \cite{Frey2011,Weber2011}\\
Raman, low resolution CRASY\saa  &      39 MHz      \cite{Lee2019}\\
Coherent anti-Stokes Raman\saa   &      30 MHz    \cite{Weber2011}\\
FTIR\replaced{\saa}{\sbb}        &      16 MHz   \cite{Albert2011}\\
Raman, high resolution CRASY\saa &     330 kHz                \scc\\
FTMW\saa                         &     few kHz \cite{Grabow2011,Shipman2011}\\
 \end{tabular}
\end{ruledtabular}\\
 \footnotesize{\saa Achieved effective resolution. \sbb Theoretical resolution limit. \scc This work.}
 \label{tab:Spectroscopic_resolution}
\end{table}%

Table \ref{tab:Spectroscopic_resolution} omitted frequency comb measurements\cite{Hansch2006,Diddams2020}\added{and related techniques}. Highest-resolution frequency comb measurement cover only very small spectral ranges and a direct comparison is therefore not meaningful. Dual-comb or direct comb spectroscopy (DCS)\cite{Foltynowicz2011,Gambetta2016,Muraviev2020} allow the rapid, broad-band, and high-resolution characterization of molecular spectra. We omit DCS from our table because it requires extended interaction times with significant molecular sample densities and, to our knowledge, the resolution of all broad-band DCS spectra is subject to significant \removed{Doppler} broadening. A comparison with estimated DCS resolution limits is therefore not meaningful \added{because the achieved effective resolution remains far below the theoretical resolution limit}.

\section{Limits to Spectroscopic Resolution and Accuracy}

The use of an infinite interferometer removes resolution limits due to the MOPD. We must therefore consider other factors that limit the resolution or accuracy of interferometric measurements. Three distinct types of uncertainties must be considered: (i) Uncertainties in delay positions accrued over the length of the opto-mechanical delay line. (ii) Uncertainties in the laser oscillator repetition rate, which affect the accuracy of the discrete 12.5 ns pulse-selection delays. (iii) Uncertainties due to Doppler shifts and Doppler broadening. In the following, we discuss each error source separately.

\subsection{Uncertainties in Opto-Mechanical Delays}

The MD-531 motorized stage in our interferometer contains a 100~nm internal encoder. The encoder is mounted on an aluminum rod with correspondingly large thermal instabilities,\footnote{The thermal expansion coefficient of aluminum at 25$^\circ$C is $1.1\cdot10^{-5} \frac{m}{mK}$ \cite{CRC_Aluminum_expansion_coefficient}.} and the calibration of known spectra revealed relative positioning errors up to  $\Delta r / r = 10^{-4}$.


We addressed the problems with the internal stage encoder by mounting an external optical encoder (Sony Laserscale BL57-RE) with a thermal expansion coefficient of $0.7 \cdot 10^{-6}$ m/(m$\cdot$K). Comparison of internal and external encoder positions revealed a very linear error for the internal encoder. The internal encoder was then calibrated over a range of 12.5 ns against the oscillator repetition rate, by measuring laser cross-correlation signals displaced by one oscillator pulse jump. With a typical $\pm 0.2\, ^{\circ}$C temperature stability on our laser table, we found that the internal encoder was sufficient to confine positioning uncertainties to $\Delta r / r < 3 \cdot 10^{-6}$ ($<40$ fs uncertainty across the stage). Higher accuracy was available by monitoring the external encoder with resulting positioning uncertainties below $\Delta r/ r < 2 \cdot 10^{-7}$ ($\approx$3 fs uncertainty across the stage).

Additional uncertainties arose due to the variation of the air refractive index $n$ with air pressure, temperature, and composition. For 800 nm light, the air refractive index changes by $\Delta n / n \approx 10^{-7}$ for a 40 Pa change in air pressure, a 0.1 $^{\circ}$C change in temperature or a 10\% change in air humidity.\cite{NIST_air_refractive_index, Ciddor1996}
These uncertainties were readily suppressed by the continuous measurement of, and correction for, changes in air temperature, pressure, and humidity. The NIST shop-floor equation\cite{NIST_air_refractive_index}\footnote{Air index of refraction $n$ based on pressure $P$ (kPa), temperature $T$ ($^{\circ}$C), and relative humidity $RH$ (\%): $n= 1+7.86 \cdot 10^{-4} \cdot P / (273 + T) - 1.5 \cdot 10^{-11} \cdot RH (T^2 + 160)$} was sufficient to approximate $n$ with a relative uncertainty of $ \Delta n/n < 10^{-7}$.

Uncertainties accrued over the range of the opto-mechanical delay line are reset with each \added{electronic} pulse selection delay jump, when the mechanical delay line is re-set to its initial position. The impact of opto-mechanical delay uncertainties therefore scales inversely to the number $N$ of \replaced{oscillator}{electronic delay} jumps and becomes small for large $N$\replaced{ and it is}{. It is therefore} sufficient to suppress relative stage positioning errors into a regime where the accrued phase shift for the highest measured frequencies becomes negligible. Stage errors in the \added{$\Delta r / r = $}$10^{-6}$ regime correspond to a 12.5 fs phase shift across the mechanically scanned 12.5 ns delay range and \replaced{are}{were} negligible for any feasible experiment with our minimal 50 fs laser pulse duration (impulsive Raman excitation possible for $<10$ THz transition frequencies). We should note that \added{when} significant calibration errors between the mechanical stage and the oscillator repetition rate \added{occurred, this} le\removed{a}d to the formation of 80 MHz sidebands\replaced{, which are}{ that were} readily identified in the experimental spectra.

\subsection{Uncertainties in the Oscillator Repetition Rate}

Extended delays are achieved by delayed electronic pulse-selection of subsequent oscillator pulses from a Coherent Vitara laser oscillator. This adds delays in multiples of 12.5 ns to the probe arm of interferometer. Any undetected drift of the laser oscillator repetition rate from the nominal value of 80 MHz will introduce a corresponding uncertainty. We used a frequency counter (Aim-TTI TF930) to monitor the stability of the oscillator against a GPS-stabilized clock (Leo Bodnar GPSDO) with an expected frequency accuracy of \added{$\Delta \nu / \nu = $}$\le10^{-10}$.

\begin{figure}[ht]
	\includegraphics[width=240pt]{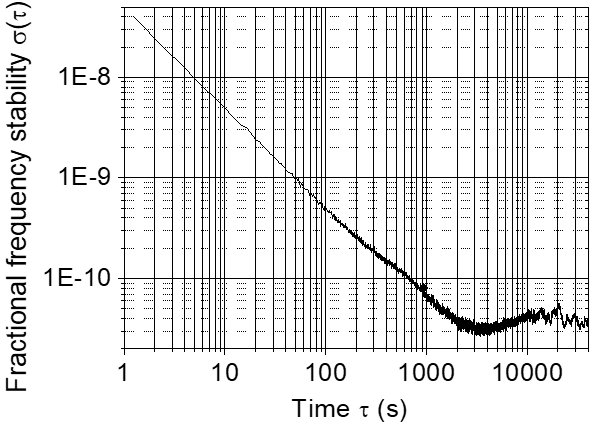}
	\caption{Allan deviation for the 80 MHz Coherent Vitara-T laser oscillator, measured against a \deleted{Leo Bodnar }GPS-disciplined clock. }
	\label{AllanDev}
\end{figure}

\begin{figure}[ht]
	\includegraphics[width=240pt]{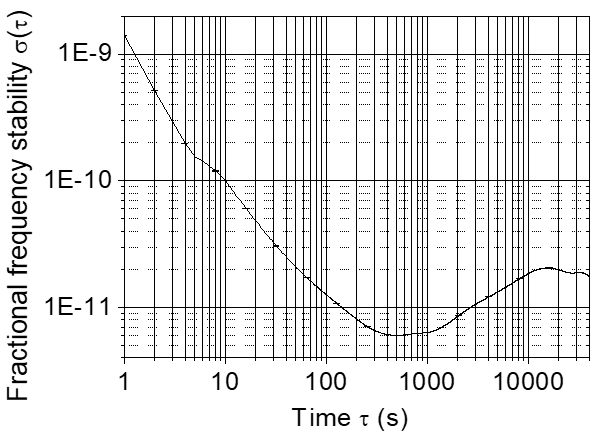}
	\caption{Modified Allan deviation for the 80 MHz Coherent Vitara-T laser oscillator, measured against a \deleted{Leo Bodnar }GPS-disciplined clock. }
	\label{AllanMDev}
\end{figure}

Figs.\ \ref{AllanDev} and \ref{AllanMDev} show the Allan deviation and the modified Allan deviation \added{for the oscillator frequency,} measured over a 1-day period. For periods $<100$ s, \replaced{we observed a}{the observed} slope of -1 (-1.5) in the Allan (modified Allan) deviation\removed{, which} is characteristic for random white noise.\cite{Riley2008} This noise is due to the frequency counter digitization noise ($\pm 1$ count over the measurement period) and does not reflect any drift of the clock or oscillator. For periods $>100$ s, the Allan deviation remained $<10^{-10}$, giving an upper limit for the frequency stability of the oscillator. It is quite possible that the GPS clock stability \replaced{is}{was} limiting in this regime and that the oscillator frequency \replaced{is}{was} more stable than our measurement indicates.


Slow frequency drifts of the oscillator are readily corrected by a corresponding adjustment of the opto-mechanical delay position. A continuous monitoring of the oscillator rate therefor allows a \emph{feed-forward} correction of delays, which fulfills the same purpose as the feed-back oscillator stabilization in frequency comb spectroscopy,\cite{Hansch2006} albeit with much smaller \replaced{technological}{technical} efforts. With our inexpensive monitoring system and a typical frequency counter period of 1 or 10 seconds, we readily achieved a single-sigma uncertainty (Allan deviation) $\Delta \nu / \nu\ll 10^{-7}$ and a longer frequency counter integration period can reduce this uncertainty \removed{down} towards the $\Delta \nu / \nu \approx 10^{-10}$ noise floor of the frequency measurement. We expect that the uncertainty can be further reduced with the use of a high-fidelity reference clock: similar oscillators \replaced{are}{were} used in frequency comb experiments and were stabilized to \replaced{several orders}{order}-of-magnitude better performance. The feedback stabilization required for the \replaced{latter}{frequency-comb measurements} introduces an additional source of noise that is absent \replaced{with}{in} our feed-forward stabilization \added{scheme}.

\subsection*{Uncertainties Arising from Doppler Effects}
\label{Doppler effects}
\begin{figure}[ht]
	\includegraphics[width=240pt]{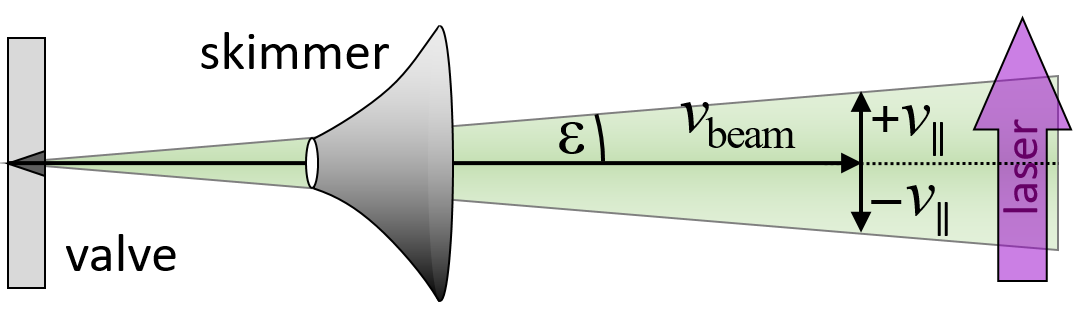}
	\caption{Illustration of Doppler broadening. The non-zero molecular beam collimation angle $\epsilon$ leads to a distribution of beam velocity components $v_\parallel$ that are parallel to the laser beam propagation axis: $v_{\parallel} = v\mathrm{_{beam} \cdot tan}(\epsilon)$.}
	\label{Doppler_broadening}
\end{figure}

\begin{figure}[ht]
	\includegraphics[width=240pt]{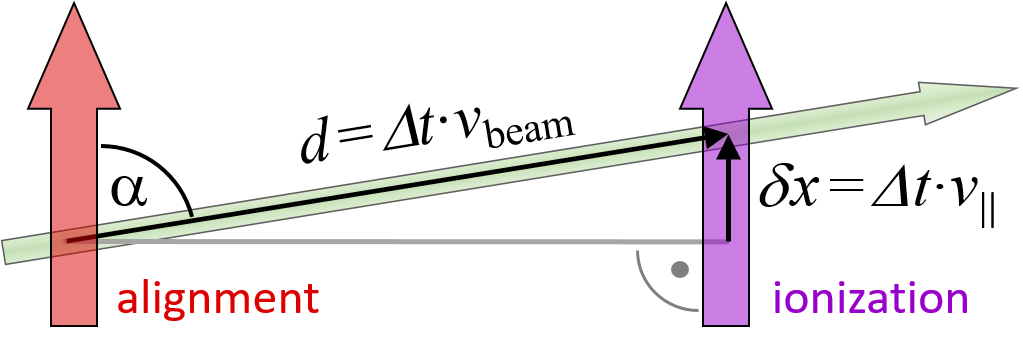}
	\caption{Illustration of the Doppler shift: To correct for molecular beam propagation within the delay $\Delta t$ between alignment and ionization pulses, the ionization laser is tracking the molecular beam for a distance of $d = \Delta t \cdot v_\mathrm{beam}$. When the angle between molecular beam and laser beams deviates from $90^{\circ}$, this leads to an additional path $\delta x$ for the ionization beam and an additional delay $\delta t = \delta x / c$.}
	\label{Doppler_shift}
\end{figure}

Textbooks commonly label skimmed molecular beam spectroscopy as "Doppler-free". But as illustrated in Figs.\ \ref{Doppler_broadening} and \ref{Doppler_shift}, the non-zero collimation angle and imperfect alignment of a molecular beam contributes some Doppler effects. We separately assessed Doppler broadening and Doppler shifts by geometric consideration of the molecular beam velocity components $v_{\parallel}$ in the direction of the alignment and ionization laser beams.

In our experiments, a 1 mm skimmer at 280 mm distance from the pulsed valve led to a molecular beam collimation angle of $\epsilon = 0.10^{\circ}$. We measured the molecular beam velocity for a helium-seeded beam of \ce{CS2} to be $v_\mathrm{beam} \approx 1100$ m/s and calculated a velocity spread $v_{\parallel} = \pm1.96$ m/s and a maximal Doppler broadening of $\Delta \nu_{\textit{Db}} / \nu = v_{\parallel} / c = 6.6 \cdot 10^{-9}$. This estimated Doppler broadening \replaced{is}{was} significantly larger than a value obtained based on the textbook treatment (see Chapter 4 in Demtr\"oder \cite{Demtroder2011}) because we accounted for the similar velocity profiles of the heavier CS$_2$ molecules and the lighter helium atoms in the seeded molecular beam.

Doppler broadening becomes relevant when it approaches or exceeds the spectroscopic resolution. In our best CRASY data, we observe sub-MHz line-width for 100 GHz line frequencies ($\Delta \nu_\mathrm{FWHM} / \nu$ in the $10^{-6}$ regime). The Doppler broadening estimated above is several orders-of-magnitude smaller than our best achieved resolution and will not affect spectroscopic results until we reach sub-kHz level resolution. The use of slower molecular beams and better molecular beam collimation can further reduce Doppler broadening and we expect that sub-100 Hz line widths \removed{for 100 GHz rotational lines} can be observed \added{for 100 GHz lines} before Doppler broadening becomes a limiting factor.

A Doppler shift occurs if the angle between laser and molecular beam deviates from $\alpha = 90^{\circ}$. As illustrated in Fig.\ \ref{Doppler_shift}, tracking of the molecular beam then changes the effective path length of the alignment versus the ionization arm of the interferometer and introduces an additional delay of $\delta t = \frac{\Delta x}{c} = \Delta t \left(1 + \frac{v_\mathrm{beam}}{c} \cdot \mathrm{sin}(\alpha) \right)$. Resulting Fourier transformed spectra show a frequency shift proportional to the delay time errors $\delta t / \Delta t$. The line position for well-resolved lines can be determined with an accuracy that is orders-of-magnitude better than the spectroscopic resolution and our experiment is therefore highly sensitive to Doppler shifts.


To measure the angle $\alpha$ between laser and molecular beam, we propagated a laser pointer through the skimmer onto the pulsed valve orifice and measured the relative angle of laser pointer beam and alignment / ionization laser beams against a reference frame. For our experiments, we determined an angle of $\alpha = 91.6 \pm 0.4 ^{\circ}$ and calculated a Doppler correction factor of $\delta t / \Delta t_{m} = (1.0 \pm0.26) \cdot 10^{-7}$. The Doppler shift was not negligible for our most-accurate measurements and reduced measured rotational frequencies by one part in $10^7$ (some 320 Hz for a 3.2 GHz rotational constant fitted for \ce{CS2}).\cite{Schroter2018} The Doppler shift uncertainty of $2.6 \cdot 10^{-8}$ can be reduced by a careful measurement of the relative angles between the molecular beam and the laser beams. The Doppler shift can be measured, and thereby completely eliminated, by performing complementary experiments with laser beams propagating in opposite directions, as measurements from opposing directions show opposite signs for the Doppler shift.

\subsection*{Signal Degradation by Sparse Sampling}

Sparse sampling is an essential tool to extend the optical path difference and thereby the spectroscopic resolution without excessive requirements in terms of measurement time and data storage. A number of sparse sampling approaches were discussed in the context of multi-dimensional NMR spectroscopy.\cite{Pelczer1991,Hoch2008,Hyberts2012,Hoch2014} Sparse sampling methods other than random sampling affect the line shape and are therefore problematic, unless the natural line shape in the investigated spectra is known or negligible. Randomly sparse sampled data merely shows an elevated noise level and correspondingly reduced signal-to-noise ratio\deleted{(SNR)}, without introducing any significant artifacts.\footnote{Random sparse sampling is equivalent to the multiplication of a continuous time-domain trace with a binary [0,1] \replaced{'white noise array'}{masking array}, which masks out the unmeasured data points. The multiplication of traces in the time domain corresponds to a folding of their spectra in the Fourier-domain. A \replaced{white noise}{random binary} array transforms into a flat \added{noise} spectrum and therefore merely adds noise \replaced{to}{without otherwise affecting} the measured spectrum.} The combination of sparse sampling with an infinite interferometer therefore offers a unique spectroscopic tool, where resolution and SNR can be freely traded against one another.

\begin{figure}[ht]
	\includegraphics[width=240pt]{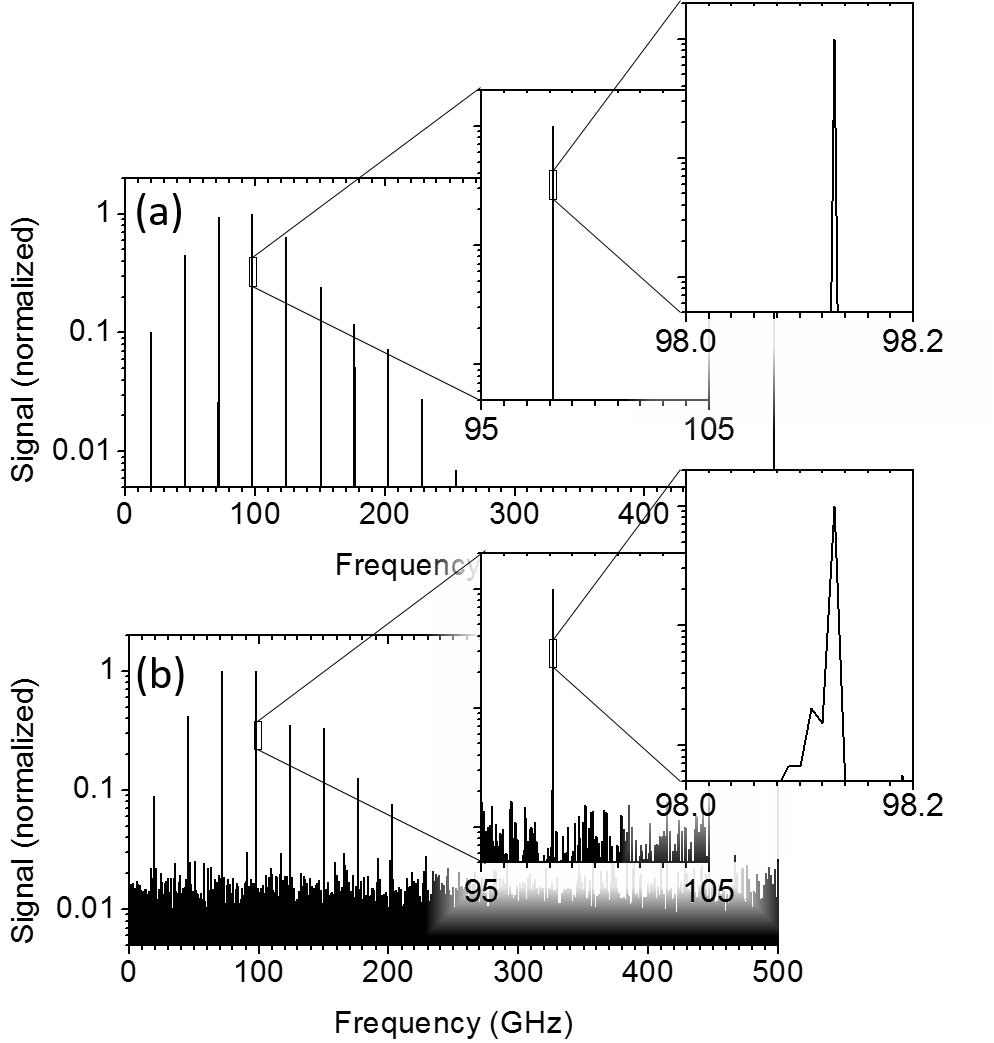}
	\caption{Simulated Fourier-transformed rotational Raman power spectrum for CS$_2$ at 8 K. (a) Full sampling of 100 ns time delay with $\Delta t$ = 1 ps steps ($10^5$ samples). (b) Sparse sampling of 3000 randomly selected delays from (a) created $< 1$\% sampling noise. Insets show a 50-fold or 2500-fold enlarged abscissa.}
	\label{Sparse_Sampling_simulation}
\end{figure}

Fig\replaced{ure}{.\ } \ref{Sparse_Sampling_simulation} shows the effect of random sparse sampling in a simulated delay trace for CS$_2$. The Fourier transform of a fully sampled trace show\replaced{s}{ed} negligible noise due to the synthetic nature of the data. To simulate 3\% random sparse sampling, 97\% of all points in the delay trace were set to zero. The random selection was based on the Mersenne-Twister pseudo-random number generator as implemented in the numpy library of the Python programming language. The Fourier transform of the sparsely sampled data show\replaced{s}{ed} significant noise but only a modest degradation of the resolution. An estimate of the noise distribution, using the modified Z-score,\cite{Iglewizc1993} gave a noise level of $\sigma$ = 0.2\% relative to the largest signal peak. Experimentally observed SNRs in sparsely sampled data showed similar signal degradation. Note that the loss of information scales proportionally to $\sqrt{\mathrm{samples}}$ and longer scans can be performed with lower sparse sampling rates.

\section{Expected Resolution Limits for CRASY Experiments}

For all practical purposes, the coherence lifetime of cold rotational states in small molecules is only limited by collisions. Rotational decoherence in collision-free, pulsed molecular beams therefore only occurs when the molecules hit the spectrometer wall. The resolution limit of CRASY is therefore purely a function of the MOPD. With the infinite interferometer design presented above, we removed all practical limitations to the MOPD and other experimental factors become limiting: (i) The molecular beam travels with supersonic velocities and must be accurately tracked. (ii) Sparse sampling is necessary to achieve large MOPDs within reasonable measurement times, but may degrade the SNR to a point where spectra can no longer be resolved.

\begin{figure}[ht]
\centering
  \includegraphics[width=240pt]{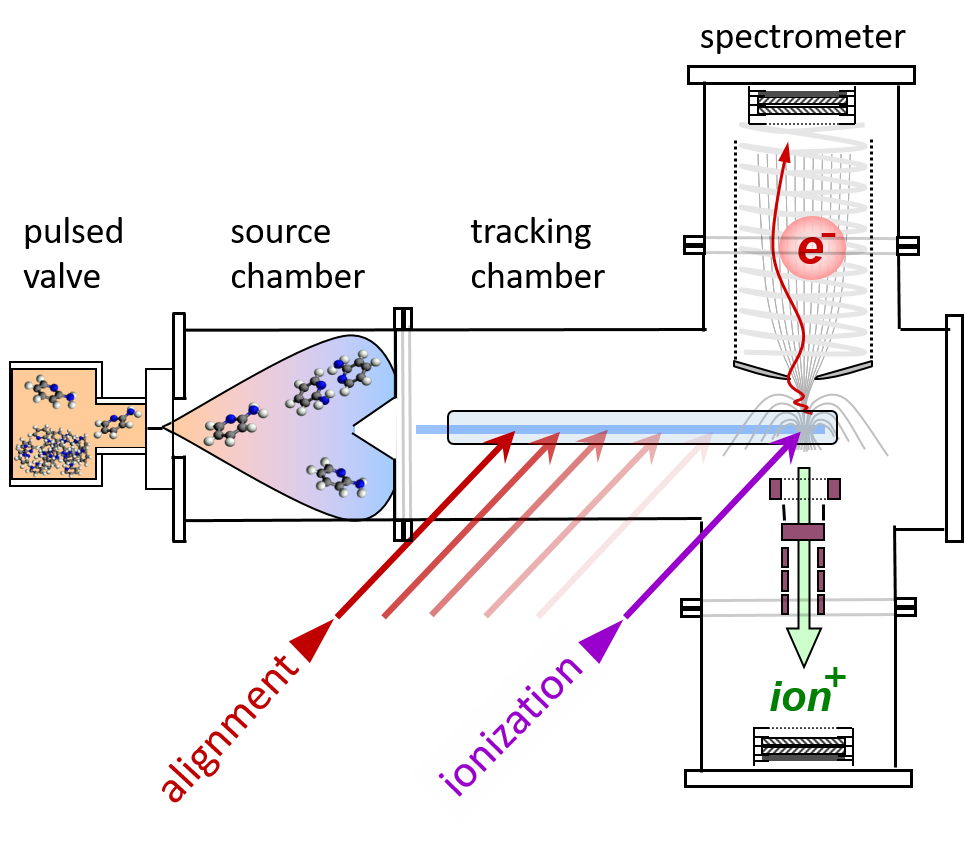}
  \caption{\small Schematic representation of a photoion-photoelectron spectrometer with extended optical access for molecular beam tracking. }
  \label{fig:Extended_tracking_chamber}
\end{figure}

The resolution of current CRASY measurements is limited by factor (i): due to a limited window size we can track the molecular beam only over distances of a few mm. The beam velocity for our dilute, helium-seeded molecular beams was measured to be $v_{b} \approx 1100$ m/s and the 330 kHz resolution data shown in Fig.\ \ref{fig:kHz_resolution_Spectrum} therefore required \removed{a} tracking of the beam over a distance of nearly 2 mm. The molecular beam velocity can be significantly reduced by using a heavier seed gas with lower speed-of-sound. With a suitably larger laser window, the tracking distance can be extended, e.g., an extension to 10 cm tracking, would offer a 50-fold increase of the accessible MOPD. We expect that the combination of longer tracking and slower molecular beams will push the resolution limit into the single-kHz regime. Further extensions would require the construction of a spectrometer with a dedicated chamber for decimeter- or meter-scale tracking of the molecular beam, e.g., as depicted in Fig.\ \ref{fig:Extended_tracking_chamber}. Note that the signal collections might be facilitated with the correlation to other spectroscopic observables. E.g., probing of rotational coherence via fluorescence excitation would remove the nonlinearity of our two-photon photoionization probe step and might allow the multiplexed detection of signals along the molecular beam axis.

\begin{figure}[ht]
\centering
  \includegraphics[width=240pt]{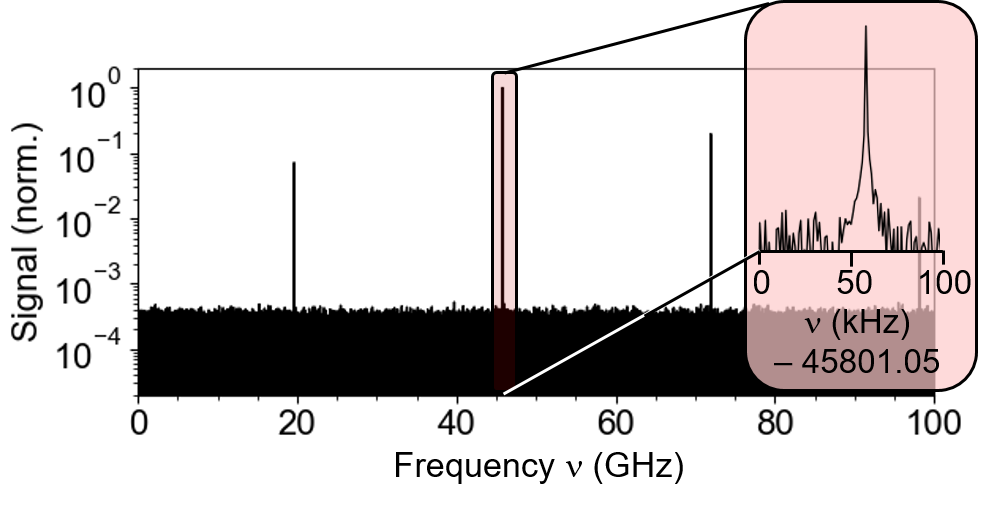}
  \caption{\small Simulation of a \ce{CS2} spectrum with 0.61 kHz non-apodized resolution, based on a 2 ms scan range, a nominal 5 ps step size, and 0.05\% sparse sampling. Simulated signal count rates were in the range of few-hundred counts per data point, corresponding to typical experimental count rates. The inset shows a 100-kHz section of the simulation.}
  \label{fig:millisecond_scan_simulation}
\end{figure}

\replaced{To collect spectroscopic data with single-kHz resolution and a 100 GHz spectral range would require the sampling of a time axis containing some $10^8$ points.}{The sampling of a time axis containing some $10^8$ points could allow to collect spectroscopic data with single-kHz resolution and a 100 GHz spectral range.} Clearly, this is only possible with severe sparse sampling and a corresponding degradation of the SNR. Fig.\ \ref{fig:millisecond_scan_simulation} shows that such measurements are feasible: a simulated spectrum based on the sampling of 100\,000 points along a 2 ms \replaced{time-delay axis}{delay range} combines excellent signal-to-noise ratio with sub-kHz resolution.

In conclusion, we demonstrated that \removed{the use of} pulse-selection from a stable laser oscillator allows to perform interferometric spectroscopy with an effectively infinite interferometer. This approach removed previous limits to the available interferometric MOPD and we presented rotational spectra with sub-MHz effective resolution over a 500 GHz spectral range range. The achieved resolution \replaced{is}{was} several orders-of-magnitude better than that achieved by any preceding RCS or FTIR measurements and corresponds to the scanning of \removed{almost} km-scale path differences. Further order-of-magnitude improvements are expected and \replaced{are}{progress is} only limited by experimental challenges such as the requirement to track skimmed molecular beams over extended distances.

\begin{acknowledgments}
The authors acknowledge funding support from the National Research Foundation of Korea, grant NRF-2018R1D1A1A02042720 and Samsung Science and Technology Foundation, grant SSTF-BA2001-08.
\end{acknowledgments}

\bibliography{References}

\begin{thebibliography}{36}%
\makeatletter
\providecommand \@ifxundefined [1]{%
 \@ifx{#1\undefined}
}%
\providecommand \@ifnum [1]{%
 \ifnum #1\expandafter \@firstoftwo
 \else \expandafter \@secondoftwo
 \fi
}%
\providecommand \@ifx [1]{%
 \ifx #1\expandafter \@firstoftwo
 \else \expandafter \@secondoftwo
 \fi
}%
\providecommand \natexlab [1]{#1}%
\providecommand \enquote  [1]{``#1''}%
\providecommand \bibnamefont  [1]{#1}%
\providecommand \bibfnamefont [1]{#1}%
\providecommand \citenamefont [1]{#1}%
\providecommand \href@noop [0]{\@secondoftwo}%
\providecommand \href [0]{\begingroup \@sanitize@url \@href}%
\providecommand \@href[1]{\@@startlink{#1}\@@href}%
\providecommand \@@href[1]{\endgroup#1\@@endlink}%
\providecommand \@sanitize@url [0]{\catcode `\\12\catcode `\$12\catcode
  `\&12\catcode `\#12\catcode `\^12\catcode `\_12\catcode `\%12\relax}%
\providecommand \@@startlink[1]{}%
\providecommand \@@endlink[0]{}%
\providecommand \url  [0]{\begingroup\@sanitize@url \@url }%
\providecommand \@url [1]{\endgroup\@href {#1}{\urlprefix }}%
\providecommand \urlprefix  [0]{URL }%
\providecommand \Eprint [0]{\href }%
\providecommand \doibase [0]{https://doi.org/}%
\providecommand \selectlanguage [0]{\@gobble}%
\providecommand \bibinfo  [0]{\@secondoftwo}%
\providecommand \bibfield  [0]{\@secondoftwo}%
\providecommand \translation [1]{[#1]}%
\providecommand \BibitemOpen [0]{}%
\providecommand \bibitemStop [0]{}%
\providecommand \bibitemNoStop [0]{.\EOS\space}%
\providecommand \EOS [0]{\spacefactor3000\relax}%
\providecommand \BibitemShut  [1]{\csname bibitem#1\endcsname}%
\let\auto@bib@innerbib\@empty
\bibitem [{\citenamefont {Schr{\"o}ter}\ \emph {et~al.}(2011)\citenamefont
  {Schr{\"o}ter}, \citenamefont {Kosma},\ and\ \citenamefont
  {Schultz}}]{Schroter2011}%
  \BibitemOpen
  \bibfield  {author} {\bibinfo {author} {\bibfnamefont {C.}~\bibnamefont
  {Schr{\"o}ter}}, \bibinfo {author} {\bibfnamefont {K.}~\bibnamefont
  {Kosma}},\ and\ \bibinfo {author} {\bibfnamefont {T.}~\bibnamefont
  {Schultz}},\ }\bibfield  {title} {\bibinfo {title} {Crasy: mass-or
  electron-correlated rotational alignment spectroscopy},\ }\href@noop {}
  {\bibfield  {journal} {\bibinfo  {journal} {Science}\ }\textbf {\bibinfo
  {volume} {333}},\ \bibinfo {pages} {1011} (\bibinfo {year}
  {2011})}\BibitemShut {NoStop}%
\bibitem [{\citenamefont {Schr{\"o}ter}\ \emph {et~al.}(2018)\citenamefont
  {Schr{\"o}ter}, \citenamefont {Lee},\ and\ \citenamefont
  {Schultz}}]{Schroter2018}%
  \BibitemOpen
  \bibfield  {author} {\bibinfo {author} {\bibfnamefont {C.}~\bibnamefont
  {Schr{\"o}ter}}, \bibinfo {author} {\bibfnamefont {J.~C.}\ \bibnamefont
  {Lee}},\ and\ \bibinfo {author} {\bibfnamefont {T.}~\bibnamefont {Schultz}},\
  }\bibfield  {title} {\bibinfo {title} {Mass-correlated rotational raman
  spectra with high resolution, broad bandwidth, and absolute frequency
  accuracy},\ }\href {https://doi.org/10.1073/pnas.1721756115} {\bibfield
  {journal} {\bibinfo  {journal} {Proceedings of the National Academy of
  Sciences}\ }\textbf {\bibinfo {volume} {115}},\ \bibinfo {pages} {5072}
  (\bibinfo {year} {2018})}\BibitemShut {NoStop}%
\bibitem [{\citenamefont {Frey}\ \emph {et~al.}(2011)\citenamefont {Frey},
  \citenamefont {Kummli}, \citenamefont {Lobsiger},\ and\ \citenamefont
  {Leutwyler}}]{Frey2011}%
  \BibitemOpen
  \bibfield  {author} {\bibinfo {author} {\bibfnamefont {H.-M.}\ \bibnamefont
  {Frey}}, \bibinfo {author} {\bibfnamefont {D.}~\bibnamefont {Kummli}},
  \bibinfo {author} {\bibfnamefont {S.}~\bibnamefont {Lobsiger}},\ and\
  \bibinfo {author} {\bibfnamefont {S.}~\bibnamefont {Leutwyler}},\ }\bibfield
  {title} {\bibinfo {title} {High-resolution rotational raman coherence
  spectroscopy with femtosecond pulses},\ }in\ \href@noop {} {\emph {\bibinfo
  {booktitle} {Handbook of High-resolution Spectroscopy}}},\ Vol.~\bibinfo
  {volume} {2},\ \bibinfo {editor} {edited by\ \bibinfo {editor} {\bibfnamefont
  {M.}~\bibnamefont {Quack}}\ and\ \bibinfo {editor} {\bibfnamefont
  {F.}~\bibnamefont {Merkt}}}\ (\bibinfo  {publisher} {John Wiley \& Sons,
  Ltd},\ \bibinfo {year} {2011})\ pp.\ \bibinfo {pages}
  {1237--1265}\BibitemShut {NoStop}%
\bibitem [{\citenamefont {Schroter}\ \emph {et~al.}(2015)\citenamefont
  {Schroter}, \citenamefont {Choi},\ and\ \citenamefont
  {Schultz}}]{Schroter2015}%
  \BibitemOpen
  \bibfield  {author} {\bibinfo {author} {\bibfnamefont {C.}~\bibnamefont
  {Schroter}}, \bibinfo {author} {\bibfnamefont {C.~M.}\ \bibnamefont {Choi}},\
  and\ \bibinfo {author} {\bibfnamefont {T.}~\bibnamefont {Schultz}},\
  }\bibfield  {title} {\bibinfo {title} {Crasy: Correlated rotational alignment
  spectroscopy reveals atomic scrambling in ionic states of butadiene},\
  }\href@noop {} {\bibfield  {journal} {\bibinfo  {journal} {Journal of
  Physical Chemistry A}\ }\textbf {\bibinfo {volume} {119}},\ \bibinfo {pages}
  {1309} (\bibinfo {year} {2015})}\BibitemShut {NoStop}%
\bibitem [{\citenamefont {Heo}\ \emph {et~al.}(2022{\natexlab{a}})\citenamefont
  {Heo}, \citenamefont {Lee}, \citenamefont {{\"O}zer},\ and\ \citenamefont
  {Schultz}}]{Heo2022a}%
  \BibitemOpen
  \bibfield  {author} {\bibinfo {author} {\bibfnamefont {I.}~\bibnamefont
  {Heo}}, \bibinfo {author} {\bibfnamefont {J.~C.}\ \bibnamefont {Lee}},
  \bibinfo {author} {\bibfnamefont {B.~R.}\ \bibnamefont {{\"O}zer}},\ and\
  \bibinfo {author} {\bibfnamefont {T.}~\bibnamefont {Schultz}},\ }\bibfield
  {title} {\bibinfo {title} {Structure of benzene from mass-correlated
  rotational raman spectroscopy},\ }\href {https://doi.org/10.1039/D2RA03431J}
  {\bibfield  {journal} {\bibinfo  {journal} {RSC Adv.}\ }\textbf {\bibinfo
  {volume} {12}},\ \bibinfo {pages} {21406} (\bibinfo {year}
  {2022}{\natexlab{a}})}\BibitemShut {NoStop}%
\bibitem [{\citenamefont {Heo}\ \emph {et~al.}(2022{\natexlab{b}})\citenamefont
  {Heo}, \citenamefont {Lee}, \citenamefont {Ã–zer},\ and\ \citenamefont
  {Schultz}}]{Heo2022b}%
  \BibitemOpen
  \bibfield  {author} {\bibinfo {author} {\bibfnamefont {I.}~\bibnamefont
  {Heo}}, \bibinfo {author} {\bibfnamefont {J.~C.}\ \bibnamefont {Lee}},
  \bibinfo {author} {\bibfnamefont {B.~R.}\ \bibnamefont {Ã–zer}},\ and\
  \bibinfo {author} {\bibfnamefont {T.}~\bibnamefont {Schultz}},\ }\bibfield
  {title} {\bibinfo {title} {Mass-correlated high-resolution spectra and the
  structure of benzene},\ }\href@noop {} {\bibfield  {journal} {\bibinfo
  {journal} {The Journal of Physical Chemistry Letters J. Phys. Chem. Lett.}\
  }\textbf {\bibinfo {volume} {13}},\ \bibinfo {pages} {8278} (\bibinfo {year}
  {2022}{\natexlab{b}})}\BibitemShut {NoStop}%
\bibitem [{\citenamefont {Grabow}(2011)}]{Grabow2011}%
  \BibitemOpen
  \bibfield  {author} {\bibinfo {author} {\bibfnamefont {J.-U.}\ \bibnamefont
  {Grabow}},\ }\bibfield  {title} {\bibinfo {title} {Fourier-transform
  microwave spectroscopy measurement and instrumentation},\ }in\ \href@noop {}
  {\emph {\bibinfo {booktitle} {Handbook of High-resolution Spectroscopy}}},\
  Vol.~\bibinfo {volume} {2},\ \bibinfo {editor} {edited by\ \bibinfo {editor}
  {\bibfnamefont {M.}~\bibnamefont {Quack}}\ and\ \bibinfo {editor}
  {\bibfnamefont {F.}~\bibnamefont {Merkt}}}\ (\bibinfo  {publisher} {John
  Wiley \& Sons, Ltd},\ \bibinfo {year} {2011})\ pp.\ \bibinfo {pages}
  {723--800}\BibitemShut {NoStop}%
\bibitem [{\citenamefont {Lee}\ \emph {et~al.}(2019)\citenamefont {Lee},
  \citenamefont {Lee},\ and\ \citenamefont {Schultz}}]{Lee2019}%
  \BibitemOpen
  \bibfield  {author} {\bibinfo {author} {\bibfnamefont {J.~C.}\ \bibnamefont
  {Lee}}, \bibinfo {author} {\bibfnamefont {D.~E.}\ \bibnamefont {Lee}},\ and\
  \bibinfo {author} {\bibfnamefont {T.}~\bibnamefont {Schultz}},\ }\bibfield
  {title} {\bibinfo {title} {High-resolution rotational raman spectroscopy of
  benzene},\ }\href {https://doi.org/10.1039/C8CP07555G} {\bibfield  {journal}
  {\bibinfo  {journal} {Phys. Chem. Chem. Phys.}\ }\textbf {\bibinfo {volume}
  {21}},\ \bibinfo {pages} {2857} (\bibinfo {year} {2019})}\BibitemShut
  {NoStop}%
\bibitem [{\citenamefont {Albert}\ \emph
  {et~al.}(2011{\natexlab{a}})\citenamefont {Albert}, \citenamefont {Albert},\
  and\ \citenamefont {Quack}}]{Albert2011}%
  \BibitemOpen
  \bibfield  {author} {\bibinfo {author} {\bibfnamefont {S.}~\bibnamefont
  {Albert}}, \bibinfo {author} {\bibfnamefont {K.~K.}\ \bibnamefont {Albert}},\
  and\ \bibinfo {author} {\bibfnamefont {M.}~\bibnamefont {Quack}},\ }\bibfield
   {title} {\bibinfo {title} {High-resolution fourier transform infrared
  spectroscopy},\ }in\ \href@noop {} {\emph {\bibinfo {booktitle} {Handbook of
  High-resolution Spectroscopy}}},\ Vol.~\bibinfo {volume} {2},\ \bibinfo
  {editor} {edited by\ \bibinfo {editor} {\bibfnamefont {M.}~\bibnamefont
  {Quack}}\ and\ \bibinfo {editor} {\bibfnamefont {F.}~\bibnamefont {Merkt}}}\
  (\bibinfo  {publisher} {John Wiley \& Sons, Ltd},\ \bibinfo {year} {2011})\
  pp.\ \bibinfo {pages} {965--1019}\BibitemShut {NoStop}%
\bibitem [{\citenamefont {Albert}\ \emph {et~al.}(2018)\citenamefont {Albert},
  \citenamefont {Bauerecker}, \citenamefont {Bekhtereva}, \citenamefont
  {Bolotova}, \citenamefont {Hollenstein}, \citenamefont {Quack},\ and\
  \citenamefont {Ulenikov}}]{Albert2018}%
  \BibitemOpen
  \bibfield  {author} {\bibinfo {author} {\bibfnamefont {S.}~\bibnamefont
  {Albert}}, \bibinfo {author} {\bibfnamefont {S.}~\bibnamefont {Bauerecker}},
  \bibinfo {author} {\bibfnamefont {E.~S.}\ \bibnamefont {Bekhtereva}},
  \bibinfo {author} {\bibfnamefont {I.~B.}\ \bibnamefont {Bolotova}}, \bibinfo
  {author} {\bibfnamefont {H.}~\bibnamefont {Hollenstein}}, \bibinfo {author}
  {\bibfnamefont {M.}~\bibnamefont {Quack}},\ and\ \bibinfo {author}
  {\bibfnamefont {O.~N.}\ \bibnamefont {Ulenikov}},\ }\bibfield  {title}
  {\bibinfo {title} {High resolution ftir spectroscopy of fluoroform
  \ce{12CHF3} and critical analysis of the infrared spectrum from 25 to 1500
  cm-1},\ }\href {https://doi.org/10.1080/00268976.2017.1392628} {\bibfield
  {journal} {\bibinfo  {journal} {Molecular Physics}\ }\textbf {\bibinfo
  {volume} {116}},\ \bibinfo {pages} {1091} (\bibinfo {year}
  {2018})}\BibitemShut {NoStop}%
\bibitem [{\citenamefont {{\"O}zer}\ \emph {et~al.}(2020)\citenamefont
  {{\"O}zer}, \citenamefont {Heo}, \citenamefont {Lee}, \citenamefont
  {Schr{\"o}ter},\ and\ \citenamefont {Schultz}}]{Ozer2020}%
  \BibitemOpen
  \bibfield  {author} {\bibinfo {author} {\bibfnamefont {B.~R.}\ \bibnamefont
  {{\"O}zer}}, \bibinfo {author} {\bibfnamefont {I.}~\bibnamefont {Heo}},
  \bibinfo {author} {\bibfnamefont {J.~C.}\ \bibnamefont {Lee}}, \bibinfo
  {author} {\bibfnamefont {C.}~\bibnamefont {Schr{\"o}ter}},\ and\ \bibinfo
  {author} {\bibfnamefont {T.}~\bibnamefont {Schultz}},\ }\bibfield  {title}
  {\bibinfo {title} {De novo structure determination of butadiene by
  isotope-resolved rotational raman spectroscopy},\ }\href
  {https://doi.org/10.1039/D0CP00129E} {\bibfield  {journal} {\bibinfo
  {journal} {Phys. Chem. Chem. Phys.}\ }\textbf {\bibinfo {volume} {22}},\
  \bibinfo {pages} {8933} (\bibinfo {year} {2020})}\BibitemShut {NoStop}%
\bibitem [{\citenamefont {Lee}\ \emph {et~al.}(2021)\citenamefont {Lee},
  \citenamefont {{\"O}zer},\ and\ \citenamefont {Schultz}}]{Lee2021}%
  \BibitemOpen
  \bibfield  {author} {\bibinfo {author} {\bibfnamefont {J.~C.}\ \bibnamefont
  {Lee}}, \bibinfo {author} {\bibfnamefont {B.~R.}\ \bibnamefont {{\"O}zer}},\
  and\ \bibinfo {author} {\bibfnamefont {T.}~\bibnamefont {Schultz}},\
  }\bibfield  {title} {\bibinfo {title} {Crasy: Correlated rotational alignment
  spectroscopy of pyridine. the rotational raman spectrum of pyridine and
  asymmetric fragmentation of pyridine dimer cation},\ }\href
  {https://doi.org/10.1039/D1CP00284H} {\bibfield  {journal} {\bibinfo
  {journal} {Phys. Chem. Chem. Phys.}\ }\textbf {\bibinfo {volume} {23}},\
  \bibinfo {pages} {10621} (\bibinfo {year} {2021})}\BibitemShut {NoStop}%
\bibitem [{\citenamefont {Stapelfeldt}(2004)}]{Stapelfeldt2004}%
  \BibitemOpen
  \bibfield  {author} {\bibinfo {author} {\bibfnamefont {H.}~\bibnamefont
  {Stapelfeldt}},\ }\bibfield  {title} {\bibinfo {title} {Laser aligned
  molecules: Applications in physics and chemistry},\ }\href@noop {} {\bibfield
   {journal} {\bibinfo  {journal} {Physica Scripta}\ }\textbf {\bibinfo
  {volume} {T110}},\ \bibinfo {pages} {132} (\bibinfo {year}
  {2004})}\BibitemShut {NoStop}%
\bibitem [{\citenamefont {Shannon}(1949)}]{Shannon1949}%
  \BibitemOpen
  \bibfield  {author} {\bibinfo {author} {\bibfnamefont {C.~E.}\ \bibnamefont
  {Shannon}},\ }\bibfield  {title} {\bibinfo {title} {Communications in the
  presence of noise.},\ }\href@noop {} {\bibfield  {journal} {\bibinfo
  {journal} {Proc. IRE}\ }\textbf {\bibinfo {volume} {37}},\ \bibinfo {pages}
  {10} (\bibinfo {year} {1949})}\BibitemShut {NoStop}%
\bibitem [{\citenamefont {Hoch}\ \emph {et~al.}(2014)\citenamefont {Hoch},
  \citenamefont {Maciejewski}, \citenamefont {Mobli}, \citenamefont
  {Schuyler},\ and\ \citenamefont {Stern}}]{Hoch2014}%
  \BibitemOpen
  \bibfield  {author} {\bibinfo {author} {\bibfnamefont {J.~C.}\ \bibnamefont
  {Hoch}}, \bibinfo {author} {\bibfnamefont {M.~W.}\ \bibnamefont
  {Maciejewski}}, \bibinfo {author} {\bibfnamefont {M.}~\bibnamefont {Mobli}},
  \bibinfo {author} {\bibfnamefont {A.~D.}\ \bibnamefont {Schuyler}},\ and\
  \bibinfo {author} {\bibfnamefont {A.~S.}\ \bibnamefont {Stern}},\ }\bibfield
  {title} {\bibinfo {title} {Nonuniform sampling and maximum entropy
  reconstruction in multidimensional nmr},\ }\href@noop {} {\bibfield
  {journal} {\bibinfo  {journal} {Accounts of Chemical Research}\ }\textbf
  {\bibinfo {volume} {47}},\ \bibinfo {pages} {708} (\bibinfo {year}
  {2014})}\BibitemShut {NoStop}%
\bibitem [{\citenamefont {Pelczer}\ and\ \citenamefont
  {Szalma}(1991)}]{Pelczer1991}%
  \BibitemOpen
  \bibfield  {author} {\bibinfo {author} {\bibfnamefont {I.}~\bibnamefont
  {Pelczer}}\ and\ \bibinfo {author} {\bibfnamefont {S.}~\bibnamefont
  {Szalma}},\ }\bibfield  {title} {\bibinfo {title} {Multidimensional nmr and
  data processing},\ }\href@noop {} {\bibfield  {journal} {\bibinfo  {journal}
  {Chemical Reviews}\ }\textbf {\bibinfo {volume} {91}},\ \bibinfo {pages}
  {1507} (\bibinfo {year} {1991})}\BibitemShut {NoStop}%
\bibitem [{\citenamefont {Albert}\ \emph {et~al.}(2015)\citenamefont {Albert},
  \citenamefont {Keppler}, \citenamefont {Lerch}, \citenamefont {Quack},\ and\
  \citenamefont {Wokaun}}]{Albert2015}%
  \BibitemOpen
  \bibfield  {author} {\bibinfo {author} {\bibfnamefont {S.}~\bibnamefont
  {Albert}}, \bibinfo {author} {\bibfnamefont {K.}~\bibnamefont {Keppler}},
  \bibinfo {author} {\bibfnamefont {P.}~\bibnamefont {Lerch}}, \bibinfo
  {author} {\bibfnamefont {M.}~\bibnamefont {Quack}},\ and\ \bibinfo {author}
  {\bibfnamefont {A.}~\bibnamefont {Wokaun}},\ }\bibfield  {title} {\bibinfo
  {title} {Synchrotron-based highest resolution ftir spectroscopy of
  chlorobenzene},\ }\href
  {https://doi.org/https://doi.org/10.1016/j.jms.2015.03.004} {\bibfield
  {journal} {\bibinfo  {journal} {Journal of Molecular Spectroscopy}\ }\textbf
  {\bibinfo {volume} {315}},\ \bibinfo {pages} {92} (\bibinfo {year} {2015})},\
  \bibinfo {note} {spectroscopy with Synchrotron Radiation}\BibitemShut
  {NoStop}%
\bibitem [{\citenamefont {Albert}\ \emph
  {et~al.}(2011{\natexlab{b}})\citenamefont {Albert}, \citenamefont {Albert},
  \citenamefont {Lerch},\ and\ \citenamefont {Quack}}]{Albert2011b}%
  \BibitemOpen
  \bibfield  {author} {\bibinfo {author} {\bibfnamefont {S.}~\bibnamefont
  {Albert}}, \bibinfo {author} {\bibfnamefont {K.~K.}\ \bibnamefont {Albert}},
  \bibinfo {author} {\bibfnamefont {P.}~\bibnamefont {Lerch}},\ and\ \bibinfo
  {author} {\bibfnamefont {M.}~\bibnamefont {Quack}},\ }\bibfield  {title}
  {\bibinfo {title} {Synchrotron-based highest resolution fourier transform
  infrared spectroscopy of naphthalene (c10h8) and indole (c8h7n) and its
  application to astrophysical problems},\ }\href
  {https://doi.org/10.1039/C0FD00013B} {\bibfield  {journal} {\bibinfo
  {journal} {Faraday Discuss.}\ }\textbf {\bibinfo {volume} {150}},\ \bibinfo
  {pages} {71} (\bibinfo {year} {2011}{\natexlab{b}})}\BibitemShut {NoStop}%
\bibitem [{\citenamefont {Shipman}\ and\ \citenamefont
  {Pate}(2011)}]{Shipman2011}%
  \BibitemOpen
  \bibfield  {author} {\bibinfo {author} {\bibfnamefont {S.~T.}\ \bibnamefont
  {Shipman}}\ and\ \bibinfo {author} {\bibfnamefont {B.~H.}\ \bibnamefont
  {Pate}},\ }\bibfield  {title} {\bibinfo {title} {New techniques in microwave
  spectroscopy},\ }in\ \href@noop {} {\emph {\bibinfo {booktitle} {Handbook of
  High-resolution Spectroscopy}}},\ Vol.~\bibinfo {volume} {2},\ \bibinfo
  {editor} {edited by\ \bibinfo {editor} {\bibfnamefont {M.}~\bibnamefont
  {Quack}}\ and\ \bibinfo {editor} {\bibfnamefont {F.}~\bibnamefont {Merkt}}}\
  (\bibinfo  {publisher} {John Wiley \& Sons, Ltd},\ \bibinfo {year} {2011})\
  pp.\ \bibinfo {pages} {801--828}\BibitemShut {NoStop}%
\bibitem [{\citenamefont {Weber}(2011)}]{Weber2011}%
  \BibitemOpen
  \bibfield  {author} {\bibinfo {author} {\bibfnamefont {A.}~\bibnamefont
  {Weber}},\ }\bibfield  {title} {\bibinfo {title} {High-resolution raman
  spectroscopy of gases},\ }in\ \href@noop {} {\emph {\bibinfo {booktitle}
  {Handbook of High-resolution Spectroscopy}}},\ Vol.~\bibinfo {volume} {2},\
  \bibinfo {editor} {edited by\ \bibinfo {editor} {\bibfnamefont
  {M.}~\bibnamefont {Quack}}\ and\ \bibinfo {editor} {\bibfnamefont
  {F.}~\bibnamefont {Merkt}}}\ (\bibinfo  {publisher} {John Wiley \& Sons,
  Ltd},\ \bibinfo {year} {2011})\ pp.\ \bibinfo {pages}
  {1153--1236}\BibitemShut {NoStop}%
\bibitem [{\citenamefont {H\"ansch}(2006)}]{Hansch2006}%
  \BibitemOpen
  \bibfield  {author} {\bibinfo {author} {\bibfnamefont {T.~W.}\ \bibnamefont
  {H\"ansch}},\ }\bibfield  {title} {\bibinfo {title} {Nobel lecture: Passion
  for precision},\ }\href@noop {} {\bibfield  {journal} {\bibinfo  {journal}
  {Reviews of Modern Physics}\ }\textbf {\bibinfo {volume} {78}},\ \bibinfo
  {pages} {1297} (\bibinfo {year} {2006})}\BibitemShut {NoStop}%
\bibitem [{\citenamefont {Diddams}\ \emph {et~al.}(2020)\citenamefont
  {Diddams}, \citenamefont {Kerry},\ and\ \citenamefont {Udem}}]{Diddams2020}%
  \BibitemOpen
  \bibfield  {author} {\bibinfo {author} {\bibfnamefont {S.~A.}\ \bibnamefont
  {Diddams}}, \bibinfo {author} {\bibfnamefont {V.}~\bibnamefont {Kerry}},\
  and\ \bibinfo {author} {\bibfnamefont {T.}~\bibnamefont {Udem}},\ }\bibfield
  {title} {\bibinfo {title} {Optical frequency combs: Coherently uniting the
  electromagnetic spectrum},\ }\href@noop {} {\bibfield  {journal} {\bibinfo
  {journal} {Science}\ }\textbf {\bibinfo {volume} {369}},\ \bibinfo {pages}
  {eaay3676} (\bibinfo {year} {2020})}\BibitemShut {NoStop}%
\bibitem [{\citenamefont {Foltynowicz}\ \emph {et~al.}(2011)\citenamefont
  {Foltynowicz}, \citenamefont {Mas{\l}owski}, \citenamefont {Ban},
  \citenamefont {Adler}, \citenamefont {Cossel}, \citenamefont {Briles},\ and\
  \citenamefont {Ye}}]{Foltynowicz2011}%
  \BibitemOpen
  \bibfield  {author} {\bibinfo {author} {\bibfnamefont {A.}~\bibnamefont
  {Foltynowicz}}, \bibinfo {author} {\bibfnamefont {P.}~\bibnamefont
  {Mas{\l}owski}}, \bibinfo {author} {\bibfnamefont {T.}~\bibnamefont {Ban}},
  \bibinfo {author} {\bibfnamefont {F.}~\bibnamefont {Adler}}, \bibinfo
  {author} {\bibfnamefont {K.~C.}\ \bibnamefont {Cossel}}, \bibinfo {author}
  {\bibfnamefont {T.~C.}\ \bibnamefont {Briles}},\ and\ \bibinfo {author}
  {\bibfnamefont {J.}~\bibnamefont {Ye}},\ }\bibfield  {title} {\bibinfo
  {title} {Optical frequency comb spectroscopy},\ }\href@noop {} {\bibfield
  {journal} {\bibinfo  {journal} {Faraday Discussions}\ }\textbf {\bibinfo
  {volume} {150}},\ \bibinfo {pages} {23} (\bibinfo {year} {2011})}\BibitemShut
  {NoStop}%
\bibitem [{\citenamefont {Gambetta}\ \emph {et~al.}(2016)\citenamefont
  {Gambetta}, \citenamefont {Cassinerio}, \citenamefont {Gatti}, \citenamefont
  {Laporta},\ and\ \citenamefont {Galzerano}}]{Gambetta2016}%
  \BibitemOpen
  \bibfield  {author} {\bibinfo {author} {\bibfnamefont {A.}~\bibnamefont
  {Gambetta}}, \bibinfo {author} {\bibfnamefont {M.}~\bibnamefont
  {Cassinerio}}, \bibinfo {author} {\bibfnamefont {D.}~\bibnamefont {Gatti}},
  \bibinfo {author} {\bibfnamefont {P.}~\bibnamefont {Laporta}},\ and\ \bibinfo
  {author} {\bibfnamefont {G.}~\bibnamefont {Galzerano}},\ }\bibfield  {title}
  {\bibinfo {title} {Scanning micro-resonator direct-comb absolute
  spectroscopy},\ }\href@noop {} {\bibfield  {journal} {\bibinfo  {journal}
  {Scientific Reports}\ }\textbf {\bibinfo {volume} {6}},\ \bibinfo {pages}
  {35541} (\bibinfo {year} {2016})}\BibitemShut {NoStop}%
\bibitem [{\citenamefont {Muraviev}\ \emph {et~al.}(2020)\citenamefont
  {Muraviev}, \citenamefont {Konnov},\ and\ \citenamefont
  {Vodopyanov}}]{Muraviev2020}%
  \BibitemOpen
  \bibfield  {author} {\bibinfo {author} {\bibfnamefont {A.~V.}\ \bibnamefont
  {Muraviev}}, \bibinfo {author} {\bibfnamefont {D.}~\bibnamefont {Konnov}},\
  and\ \bibinfo {author} {\bibfnamefont {K.~L.}\ \bibnamefont {Vodopyanov}},\
  }\bibfield  {title} {\bibinfo {title} {Broadband high-resolution molecular
  spectroscopy with interleaved mid-infrared frequency combs},\ }\href@noop {}
  {\bibfield  {journal} {\bibinfo  {journal} {Scientific Reports}\ }\textbf
  {\bibinfo {volume} {10}},\ \bibinfo {pages} {18700} (\bibinfo {year}
  {2020})}\BibitemShut {NoStop}%
\bibitem [{Note1()}]{Note1}%
  \BibitemOpen
  \bibinfo {note} {The thermal expansion coefficient of aluminum at 25$^\circ
  $C is $1.1\cdot 10^{-5} \protect \frac {m}{mK}$ \cite
  {CRC_Aluminum_expansion_coefficient}.}\BibitemShut {Stop}%
\bibitem [{\citenamefont {Stone~Jr.}\ and\ \citenamefont
  {Zimmerman}(2001)}]{NIST_air_refractive_index}%
  \BibitemOpen
  \bibfield  {author} {\bibinfo {author} {\bibfnamefont {J.~A.}\ \bibnamefont
  {Stone~Jr.}}\ and\ \bibinfo {author} {\bibfnamefont {J.~H.}\ \bibnamefont
  {Zimmerman}},\ }\href
  {https://www.nist.gov/publications/index-refraction-air} {\bibinfo {title}
  {Index of refraction of air}},\ \bibinfo {howpublished}
  {\url{https://www.nist.gov/publications/index-refraction-air}} (\bibinfo
  {year} {2001}),\ \bibinfo {note} {accessed: 2022-09-22}\BibitemShut {NoStop}%
\bibitem [{\citenamefont {Ciddor}(1996)}]{Ciddor1996}%
  \BibitemOpen
  \bibfield  {author} {\bibinfo {author} {\bibfnamefont {P.~E.}\ \bibnamefont
  {Ciddor}},\ }\bibfield  {title} {\bibinfo {title} {Refractive index of air:
  new equations for the visible and near infrared},\ }\href@noop {} {\bibfield
  {journal} {\bibinfo  {journal} {Applied Optics}\ }\textbf {\bibinfo {volume}
  {35}},\ \bibinfo {pages} {1566} (\bibinfo {year} {1996})}\BibitemShut
  {NoStop}%
\bibitem [{Note2()}]{Note2}%
  \BibitemOpen
  \bibinfo {note} {Air index of refraction $n$ based on pressure $P$ (kPa),
  temperature $T$ ($^{\circ }$C), and relative humidity $RH$ (\%): $n= 1+7.86
  \cdot 10^{-4} \cdot P / (273 + T) - 1.5 \cdot 10^{-11} \cdot RH (T^2 +
  160)$}\BibitemShut {NoStop}%
\bibitem [{\citenamefont {Riley}(2008)}]{Riley2008}%
  \BibitemOpen
  \bibfield  {author} {\bibinfo {author} {\bibfnamefont {W.}~\bibnamefont
  {Riley}},\ }\href@noop {} {\emph {\bibinfo {title} {Handbook of Frequency
  Stability Analysis}}},\ National Institute of Standards and Technology
  Special Publication 1065\ (\bibinfo  {publisher} {U. S. GOVERNMENT PRINTING
  OFFICE},\ \bibinfo {address} {Washington},\ \bibinfo {year}
  {2008})\BibitemShut {NoStop}%
\bibitem [{\citenamefont {Demtroder}(2011)}]{Demtroder2011}%
  \BibitemOpen
  \bibfield  {author} {\bibinfo {author} {\bibfnamefont {W.}~\bibnamefont
  {Demtroder}},\ }\bibfield  {title} {\bibinfo {title} {Doppler-free laser
  spectroscopy},\ }in\ \href@noop {} {\emph {\bibinfo {booktitle} {Handbook of
  High-resolution Spectroscopy}}},\ Vol.~\bibinfo {volume} {3},\ \bibinfo
  {editor} {edited by\ \bibinfo {editor} {\bibfnamefont {M.}~\bibnamefont
  {Quack}}\ and\ \bibinfo {editor} {\bibfnamefont {F.}~\bibnamefont {Merkt}}}\
  (\bibinfo  {publisher} {John Wiley \& Sons, Ltd},\ \bibinfo {year} {2011})\
  pp.\ \bibinfo {pages} {1759--1779}\BibitemShut {NoStop}%
\bibitem [{\citenamefont {Hoch}\ \emph {et~al.}(2008)\citenamefont {Hoch},
  \citenamefont {Maciejewski},\ and\ \citenamefont {Filipovic}}]{Hoch2008}%
  \BibitemOpen
  \bibfield  {author} {\bibinfo {author} {\bibfnamefont {J.~C.}\ \bibnamefont
  {Hoch}}, \bibinfo {author} {\bibfnamefont {M.~W.}\ \bibnamefont
  {Maciejewski}},\ and\ \bibinfo {author} {\bibfnamefont {B.}~\bibnamefont
  {Filipovic}},\ }\bibfield  {title} {\bibinfo {title} {Randomization improves
  sparse sampling in multidimensional nmr},\ }\href@noop {} {\bibfield
  {journal} {\bibinfo  {journal} {Journal of magnetic resonance (San Diego,
  Calif. : 1997)}\ }\textbf {\bibinfo {volume} {193}},\ \bibinfo {pages} {317}
  (\bibinfo {year} {2008})}\BibitemShut {NoStop}%
\bibitem [{\citenamefont {Hyberts}\ \emph {et~al.}(2012)\citenamefont
  {Hyberts}, \citenamefont {Milbradt}, \citenamefont {Wagner}, \citenamefont
  {Arthanari},\ and\ \citenamefont {Wagner}}]{Hyberts2012}%
  \BibitemOpen
  \bibfield  {author} {\bibinfo {author} {\bibfnamefont {S.~G.}\ \bibnamefont
  {Hyberts}}, \bibinfo {author} {\bibfnamefont {A.~G.}\ \bibnamefont
  {Milbradt}}, \bibinfo {author} {\bibfnamefont {A.~B.}\ \bibnamefont
  {Wagner}}, \bibinfo {author} {\bibfnamefont {H.}~\bibnamefont {Arthanari}},\
  and\ \bibinfo {author} {\bibfnamefont {G.}~\bibnamefont {Wagner}},\
  }\bibfield  {title} {\bibinfo {title} {Application of iterative soft
  thresholding for fast reconstruction of nmr data non-uniformly sampled with
  multidimensional poisson gap scheduling},\ }\href@noop {} {\bibfield
  {journal} {\bibinfo  {journal} {Journal of Biomolecular NMR}\ }\textbf
  {\bibinfo {volume} {52}},\ \bibinfo {pages} {315} (\bibinfo {year}
  {2012})}\BibitemShut {NoStop}%
\bibitem [{Note3()}]{Note3}%
  \BibitemOpen
  \bibinfo {note} {Random sparse sampling is equivalent to the multiplication
  of a continuous time-domain trace with a binary [0,1] masking array, which
  masks out the unmeasured data points. The multiplication of traces in the
  time domain corresponds to a folding of their spectra in the Fourier-domain.
  A random binary array transforms into a flat {noise} spectrum and therefore
  merely adds noise without otherwise affecting the measured
  spectrum.}\BibitemShut {Stop}%
\bibitem [{\citenamefont {Iglewicz}\ and\ \citenamefont
  {Hoaglin}(1993)}]{Iglewizc1993}%
  \BibitemOpen
  \bibfield  {author} {\bibinfo {author} {\bibfnamefont {B.}~\bibnamefont
  {Iglewicz}}\ and\ \bibinfo {author} {\bibfnamefont {D.}~\bibnamefont
  {Hoaglin}},\ }\bibfield  {title} {\bibinfo {title} {How to detect and handle
  outliers},\ }in\ \href@noop {} {\emph {\bibinfo {booktitle} {The ASQC Basic
  References in Quality Control: Statistical Techniques}}},\ Vol.~\bibinfo
  {volume} {16},\ \bibinfo {editor} {edited by\ \bibinfo {editor}
  {\bibfnamefont {E.}~\bibnamefont {Mykytka}}}\ (\bibinfo  {publisher}
  {American Society for Quality Control},\ \bibinfo {year} {1993})\ \bibinfo
  {note} {see also: NIST/SEMATECH e-Handbook of Statistical Methods,
  http://www.itl.nist.gov/div898/handbook/, accessed Feb. 15, 2017}\BibitemShut
  {NoStop}%
\bibitem [{\citenamefont {Lide}(2005)}]{CRC_Aluminum_expansion_coefficient}%
  \BibitemOpen
  \bibfield  {author} {\bibinfo {author} {\bibfnamefont {D.~E.}\ \bibnamefont
  {Lide}},\ }\bibinfo {title} {Crc handbook of chemistry and physics}\
  (\bibinfo  {publisher} {CRC press},\ \bibinfo {address} {Boca Raton},\
  \bibinfo {year} {2005})\ pp.\ \bibinfo {pages} {12--196},\ \bibinfo {edition}
  {86th}\ ed.\BibitemShut {Stop}%
\end{thebibliography}%

\end{document}